\documentclass[%
 aip,
 amsmath,amssymb,
 reprint,%
]{revtex4-1}

\usepackage{graphicx}
\usepackage{dcolumn}
\usepackage{bm}

\usepackage[utf8]{inputenc}
\usepackage[T1]{fontenc}
\usepackage{mathptmx}
\usepackage{etoolbox}

\usepackage{graphicx}
\usepackage{float}
\usepackage{subfigure}
\usepackage{amsmath}
\usepackage{textcomp}
\usepackage{lipsum}
\usepackage{xcolor}
\usepackage{appendix}
\begin{document}
\title{Long lasting plasma density structures utilizing tailored density profiles}
\author{M.~Luo}\email{mufei.luo@physics.ox.ac.uk}
\address{Department of Physics, Chalmers University of Technology,  G\"{o}teborg, SE-41296, Sweden}
\address{Department of Physics, University of Oxford, Parks Road, Oxford OX1 3PU, UK}
\author{C.~Riconda} 
\address{LULI, Sorbonne Université, CNRS, École Polytechnique, CEA, 75252 Paris, France}
\author{A.~Grassi}
\address{LULI, Sorbonne Université, CNRS, École Polytechnique, CEA, 75252 Paris, France}
\author{N.~Wang}
\address{College of Electrical Engineering, Zhejiang University, Hangzhou 310027, China}
\author{J.~S.~Wurtele}
\address{Department of Physics, University of California, Berkeley, California 94720, USA}
\author{I.~Pusztai}
\address{Department of Physics, Chalmers University of Technology,  G\"{o}teborg, SE-41296, Sweden}
\author{T.~F\"ul\"op}
\address{Department of Physics, Chalmers University of Technology,  G\"{o}teborg, SE-41296, Sweden}
%
\begin{abstract}
Using fully kinetic Particle-In-Cell simulations, we investigate the stability and performance of autoresonant plasma beat–wave excitation in plasmas with tailored density profiles. We show that a prescribed spatial variation of the background density sustains continuous phase locking between the driving laser beat and the excited plasma mode, thereby enabling precise control of the plasma-wave packet’s shape and group velocity and providing an alternative to frequency chirping of the drive lasers. The density-gradient scale is found to govern the nonlinear autoresonant growth, and the attainable saturation amplitude can exceed the classical Rosenbluth-Liu prediction and, for appropriate laser intensities, approach the nonrelativistic wave-breaking limit.   We show that a four-laser configuration in a steep parabolic density profile can generate a specially confined two-phase quasi-periodic plasma lattice. The generation of such structures may lead to novel applications in plasma photonics.
\end{abstract}
\maketitle

\date{today} 

\section{Introduction \label{introduction}}

The plasma beat-wave accelerator (PBWA), originally proposed by Tajima and Dawson~\cite{Tajimaprl1979}, offers a compelling mechanism for generating relativistic plasma waves via the ponderomotive force of two co-propagating laser pulses. Unlike laser wakefield acceleration (LWFA)—which employs a single, ultrashort femtosecond pulse~\cite{leemans2006gev,oubrerie2022controlled,PhysRevLett.128.164801}—PBWA utilizes multi-picosecond laser pulses, providing greater flexibility in choosing plasma and laser parameters. This includes relaxed constraints on laser diffraction~\cite{PhysRevAccelBeams.26.061301}, enhanced electron trapping near critical densities~\cite{photonics9070476}, and the possibility of tailoring the plasma wave's phase velocity via engineered plasma channels~\cite{plasma6010003}. In addition to particle acceleration, PBWA has emerged as a promising source of intense terahertz (THz) radiation, generated through mechanisms such as linear mode conversion~\cite{linearmode3}, transient current generation~\cite{current1}, and coherent plasma oscillations~\cite{minsup}. Moreover, two-phase beat-wave drives can produce quasi-crystalline plasma structures, opening new directions in plasma photonics~\cite{Munirovpre,Munirovprr}.

In conventional PBWA, the plasma wave amplitude is limited by detuning effects associated with nonlinear wavelength shifts. These shifts cause a breakdown of resonant coupling and ultimately limit the electric field to the Rosenbluth–Liu (RL) limit~\cite{Rosenbluthprl1972}, $E_{\rm RL} = (16 a_1 a_2 / 3)^{1/3} E_0$, where $E_0 = m_e c \omega_{\rm pe}/e$ is the nonrelativistic cold wave-breaking field. Here, $a_{1,2}$ are the normalized vector potentials of the lasers, defined as $a_{1,2} = eA_{1,2} / m_e c$, and $\omega_{\rm pe} = \sqrt{n_e e^2 / \epsilon_0 m_e}$ is the electron plasma frequency. 

To overcome this nonlinear saturation, autoresonance, a powerful nonlinearity control mechanism~\cite{Fajansajp2001}, has been applied. Previous studies demonstrated that applying a controlled frequency chirp to one of the laser beams can maintain phase-locking and drive the wave beyond the RL limit~\cite{lindbergprl,lindbergpop,Meerson,luoprr,luojpp}. Autoresonance has also been explored in other plasma contexts, including inertial confinement fusion (ICF), where it influences the growth of unwanted instabilities such as stimulated Raman scattering (SRS)~\cite{Chapmanprl,Yaakobi2008,Luopop} and stimulated Brillouin scattering (SBS)~\cite{stefan1991,stefan2} in tailored density or flow profiles.

In this work, we investigate a chirp-free, density-driven autoresonant plasma wave using fully kinetic Particle-In-Cell (PIC) simulations. We show that spatially tailored plasma density profiles alone can achieve continuous phase-locking, enabling the plasma wave to grow to amplitudes near the wave-breaking limit. We examine how the plasma wave's amplitude, shape, and propagation characteristics depend on laser intensity and plasma density gradients. Systematic studies are conducted for both linearly increasing and parabolic density profiles, and scaling laws guided by theoretical considerations are compared with simulation results.
While in the first part of the paper we focus on traveling waves, autoresonant excitation can also be used to generate standing  electron plasma waves \cite{Friedland_Shagalov_2020} and ion acoustic waves \cite{frieland2019}, or more generally two-phase solutions \cite{Friedland_Shagalov_2020,Munirovpre,Munirovprr}. 
In the above mentioned papers, the autoresonant mechanism was triggered by two counter-propagating ponderomotive drives with a chirped frequency difference. Here we show that spatial, density-driven autoresonance can also generate a large amplitude two-phase electron plasma wave that can be used as a quasi-crystal.

The remainder of the paper is organized as follows: 
autoresonant excitation in linear and parabolic density profiles using one pair of laser beams is analyzed in Sec.~\ref{parabolic-part}. Two counter-propagating pairs of laser beams are then utilized to create a quasi-crystalline plasma wave structure described in Sec.~\ref{crystal}. Finally, we summarize and discuss implications in Sec.~\ref{conclusion}.

\section{Autoresonant plasma wave excitation in a varying density profile \label{parabolic-part}}

A plasma wave excited by PWBA \textcolor{black}{in a varying density profile $n_e(x)/n_{\rm re}$ (here $n_{\rm re}$ denotes the reference density, which is discussed in the following)} will experience a nonlinear wavelength shift that, in the weakly nonlinear regime ($E_L/\tilde{E}_0 \leq 1$, with the longitudinal electrostatic field $E_L$), is given by \cite{RevModPhys.81.1229}
\begin{equation}
\begin{aligned}
\lambda_{\rm np}
&= \lambda_{\rm p}(1+3(E_L/\tilde{E}_0)^2/16) \\
&= \lambda_{\rm p}(1+3(E_L/E_0)^2(n_{\rm re}/n_e)/16)
\end{aligned}
\end{equation}
where $\lambda_{\rm p} = 2\pi/k_{\rm pe} = 2\pi c/\omega_{\rm pe}$ is the linear plasma wavelength \textcolor{black}{at the density $n_e(x)$, while $\tilde{E}_0(x)$ and $E_0$ are the wave-breaking limits for density $n_e(x)$ and $n_{\rm re}$, respectively.} Correspondingly, the nonlinear wavenumber becomes $k_{\rm np}=2\pi/\lambda_{\rm np}= k_{\rm re}+\delta k_{\rm non}$, with the relative shift 
\begin{equation}
\delta k_{\rm non}/k_{\rm re}=\frac{\sqrt{n_e(x)/n_{\rm re}}}{1+3(E_L/E_0)^2(n_{\rm re}/n_e(x))/16}-1,    
\label{nonlinear_wavenumber}
\end{equation}
\textcolor{black}{here $k_{\rm re}$ denotes the wavenumber for density $n_{\rm re}$.} This nonlinear shift introduces a phase mismatch between the laser beat and the plasma wave that  limits the wave amplitude. Specifically, the plasma wave saturates at the RL limit~\cite{Rosenbluthprl1972}, $E_L/\tilde{E}_0 = (16 a_1 a_2 / 3)^{1/3}$, beyond which the three-wave coupling is effectively detuned due to the nonlinear frequency shift.

To overcome this amplitude limitation, a frequency chirp can be applied to one of the laser beams, enabling the plasma autoresonant wave amplitude to exceed the RL limit and potentially approach the nonrelativistic wave-breaking field. This idea was investigated in fluid models~\cite{lindbergpop,lindbergprl} and with PIC simulations~\cite{luoprr,luojpp}. Alternatively, a tailored plasma density profile can be employed~\cite{Chapmanprl,Luobandwidth,Luopop}. In this case, the spatial variation of the plasma frequency compensates for the nonlinear detuning, allowing controlled excitation of the plasma wave. In the following, we investigate, with PIC simulations, autoresonant large-amplitude plasma wave generation in linearly increasing and parabolic density profiles and compare our results with qualitative theoretical expectations.

In an inhomogeneous plasma, the local plasma wavenumber evolves due to both spatial density variation and nonlinear wave dynamics. Without nonlinearities it can be expressed as
\begin{equation}
  k_{\rm pe}=k_{\rm re}+\delta k_s, 
  \label{local wavenumer}
\end{equation}
where we define $\delta k_{\rm s}$, the wavenumber shift related to the spatially varying background plasma density. The plasma density and corresponding local plasma frequency are then defined as
\begin{align}
    n_e(x) &= n_{\rm re} \left[1 + \left(\frac{x - x_{\rm re}}{L_{\rm gra}}\right)^i \right], \label{parabolic-density} \\
    \omega_{\rm pe}^2(x) &= \omega_{\rm re}^2 \left[1 + \left(\frac{x - x_{\rm re}}{L_{\rm gra}}\right)^i \right], \label{parabolic-frequency} \\
    k_{\rm pe}(x)&=\omega_{\rm pe}(x)/c \label{parabolic-full-wavenumber}.
\end{align}
Here the exponent $i=1$ refers to the linear profile and $i=2$ refers to the parabolic profile.  
Equations~\eqref{parabolic-frequency} and~\eqref{parabolic-full-wavenumber} define the local plasma frequency and the local linear resonance condition for the plasma wave \textcolor{black}{in the cold plasma limit (kinetic corrections are neglected)}, respectively, and $L_{\rm gra}$ denotes the characteristic gradient length of the plasma density. Note that in the parabolic case  the spatial variation of the local plasma frequency is symmetric around the linear resonant point $x_{\rm re}$. In that point $k_{\rm pe}(x=x_{\rm re})=k_{\rm re}$ and, henceforth, $k_{\rm re}^{-1}$ is used to normalize length, while time is normalized to $\omega_{\rm re}^{-1}$.

The density-induced wavenumber shift is 
\begin{equation}
    \delta k_{\rm s}(x)/k_{\rm re} = \sqrt{n_e(x)/n_{\rm re}}-1. \label{parabolic-wavenumber}
\end{equation}
Including nonlinear effects and assuming perfect compensation, we have 
\begin{equation}
   \delta k_s+\delta k_{\rm non}=0. \label{condition}
\end{equation}
Recalling Eq.~(\ref{nonlinear_wavenumber}), with $E_0$ evaluated at $n_{\rm re}$, the plasma wave amplitude can be estimated by
\textcolor{black}{
\begin{equation}
    \left(\frac{E_L}{E_0}\right)^2 = \frac{16}{3}\frac{n_e(x)}{n_{\rm re}}\left(\frac{\sqrt{n_e(x)/n_{\rm re}}}{2-\sqrt{n_e(x)/n_{\rm re}}}-1\right). \label{el}
\end{equation}}
%
\textcolor{black}{We note that the mathematical possibility of a divergence in Eq.~(\ref{el}) when $n_e/n_{\rm re}=4$ is not a concern, since $n_e/n_{\rm re}$ remains close to unity, as we shall see later (e.g.~in Fig.~\ref{scaling}).}

In the following we present numerical results characterizing the autoresonant plasma wave excitation process in a plasma with a linearly increasing (Sec.~\ref{linear part1}) and parabolic (Sec.~\ref{parabolic part1}) density profiles, respectively.
\begin{figure*}[htbp]
    \centering
    \includegraphics[width=1\linewidth]{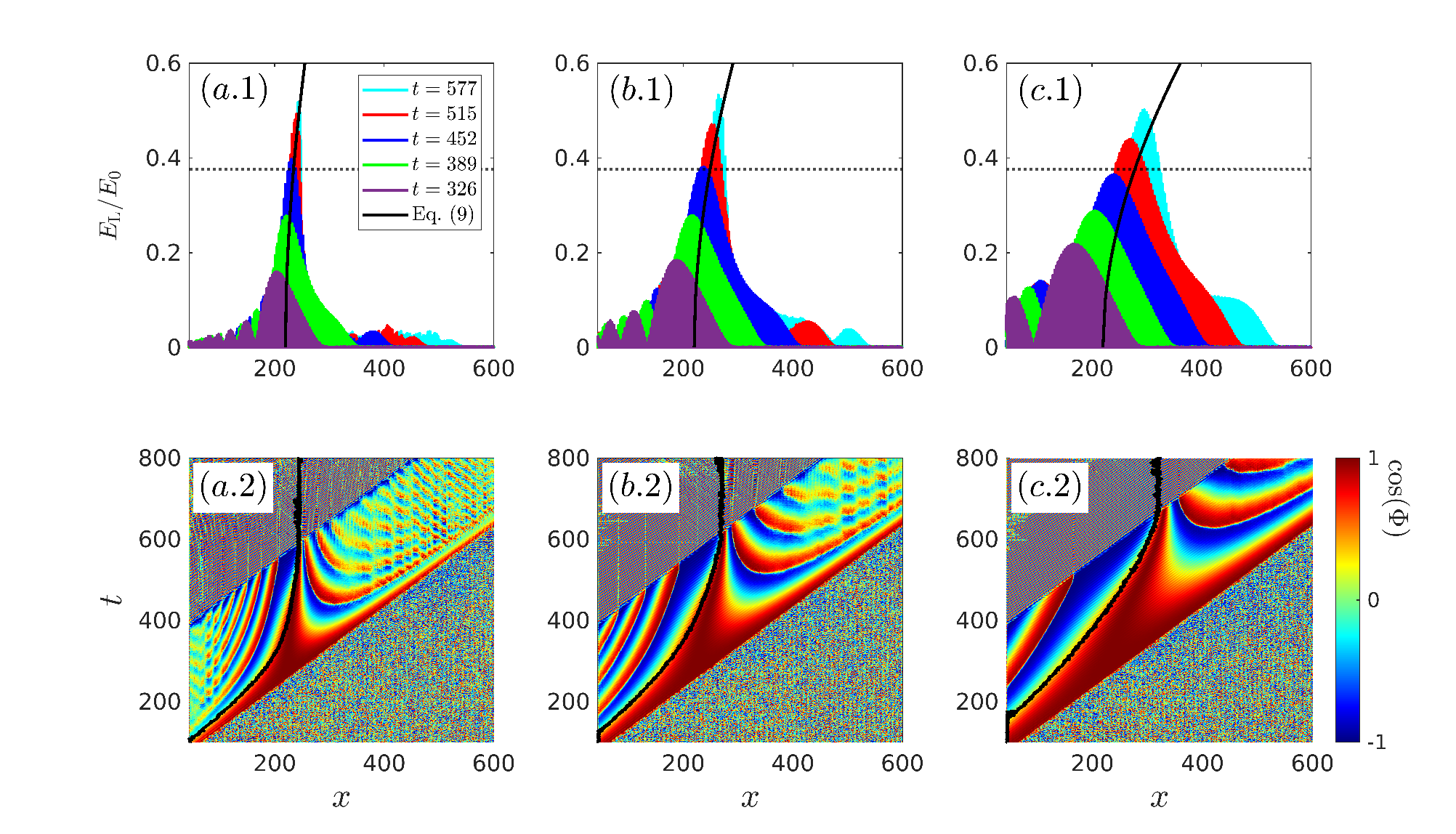}
    \caption{Top panels: The spatial profile of the plasma wave at different times, for the linearly increasing plasma density profile. Solid black curves correspond to the amplitude prediction Eq.~(\ref{el}), and the dotted black line indicates the RL limit at the reference point. Bottom panels: The spatio-temporal evolution of the phase difference between the three interacting waves: $\Phi(x)=\phi_1(x)-\phi_2(x)-\phi_L(x)$, with $\phi_{1}$, $\phi_{2}$ and $\phi_{L}$ representing the complex phase of the laser beams and the plasma wave, respectively. Here, the solid black curves mark the location of the highest plasma wave amplitude. The density gradient length is increased between the panels from left to right: $L=180\pi$, $L=360\pi$, and $L=720\pi$. The simulations use the normalized laser amplitude $a_1=a_2=0.1$ and the reference plasma density $n_{\rm re}(x_{\rm re})=0.0004n_{\rm cr}$.}    
\label{90_180_360linear} 
\end{figure*}

\subsection{Autoresonant plasma wave excitation in a linearly increasing plasma density \label{linear part1}}

\textcolor{black}{To study the autoresonant plasma wave excitation mechanism, we performed one-dimensional (1D) fully kinetic simulations using the {\sc Smilei} PIC code~\cite{DEROUILLAT2018351}. The simulation involves two co-propagating laser beams with identical intensities and parallel linear polarizations. Each pulse has a duration of $T_0 = 100\pi/\omega_{\rm re}$, where $\omega_{\rm re} = \omega_{\rm pe}(x_{\rm re})$ is the electron plasma frequency evaluated at the reference position $x_{\rm re} = 70\pi$ (the left boundary has $x=0$), and $\omega_{\rm re}$ is precisely matched to the frequency difference between the two laser beams, as discussed in the following. The pulse duration was chosen to be sufficiently short to prevent the development of ion-related instabilities during the interaction~\cite{Mora1988prl}.}

\textcolor{black}{The laser envelopes follow a 16th-order super-Gaussian longitudinal profile. The electron density at the reference point is set to $n_{\rm re} = 0.0004\,n_{\rm cr}$, corresponding to an extremely underdense plasma that minimizes nonlinear optical effects such as Raman scattering and frequency shifts~\cite{luoprr}.} \textcolor{black}{The laser frequencies satisfy the beat resonance condition $\omega_1 - \omega_2 = \omega_{\rm re}$, with $\omega_1 = 50\,\omega_{\rm re}$ and $\omega_2 = 49\,\omega_{\rm re}$. Vacuum buffers of length $L_{\rm v} = 14\pi$ are inserted at both ends of the domain to suppress boundary effects. Ions are modeled as immobile, a valid approximation for the timescale of the interaction~\cite{Mora1988prl}, this was also verified by additional simulations with mobile ions (not shown). 
We normalize distance by $k_{\rm re}^{-1}$, the reference plasma wavenumber, and time by $\omega_{\rm re}^{-1}$, the reference plasma frequency. 
The spatial resolution is set to ${\rm d}x = 0.008$, with a time step of $c\,{\rm d}t = 0.9\,{\rm d}x$, satisfying the Courant–Friedrichs–Lewy stability criterion. Each cell is initialized with 50 macro-particles per species. The simulations do not include Coulomb collisions.}
\begin{figure*}[htbp]
    \centering
    \includegraphics[width=1\linewidth]{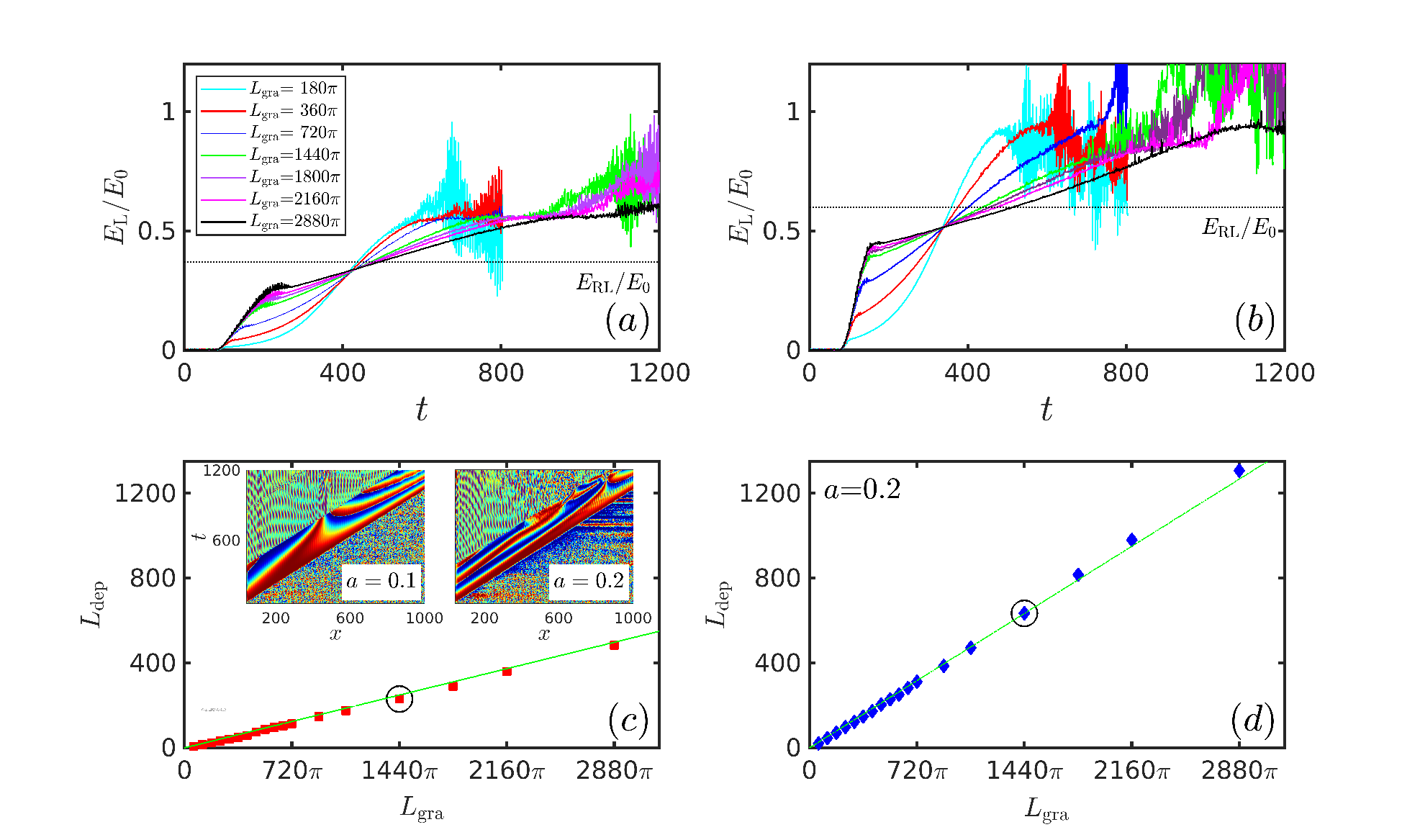}
    \caption{Top panels: The maximum value of $E_L/E_0$ over the simulation domain as a function of time for a wide range of gradient lengths (see legend in (a)); The horizontal black dotted represents the RL limit at the reference point $x_{\rm re}$. Bottom panels: Dephasing length $L_{\rm dep}$ as a function of density gradient length. The laser amplitude for (a) and (c) is $a_{1,2}$=0.1 and for (b) and (d) $a_{1,2}$=0.2. The inserts in panel (c) show the spatio-temporal evolution of the phase difference $\Phi(x)$ at a fixed gradient length $L_{\rm gra}=1440\pi$, for intensities $a_{1,2}=0.1$ and $a_{1,2}=0.2$; these cases are indicated by the open black circles in panel (c) and (d). The reference density is $n_{\rm re}(x_{\rm re})=0.0004n_{\rm cr}$. The green lines in panel (c) and (d) are predictions using Eq.~(\ref{linear dephase1}).}  
\label{a0.10.2} 
\end{figure*}
\begin{figure}[htbp]
    \centering
    \includegraphics[width=1\linewidth]{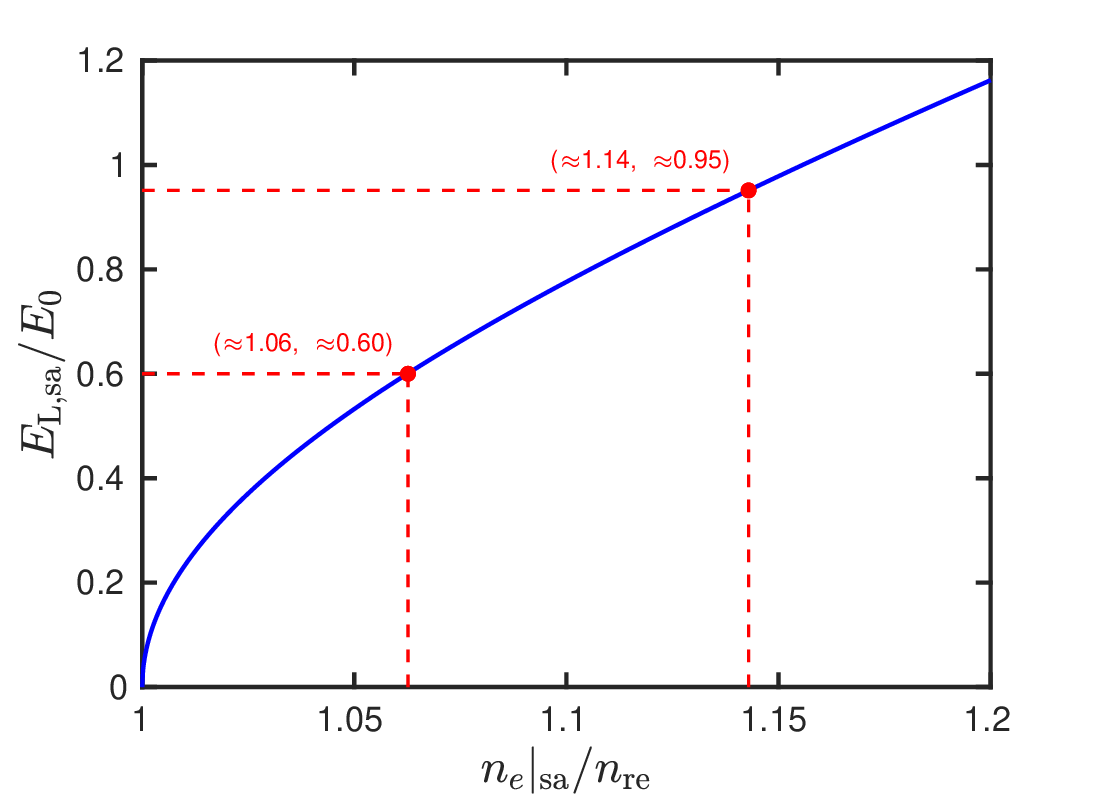}
    \caption{\textcolor{black}{The relation between \(n_e|_{\rm sa}/n_{\rm re}\) and the electric field \(E_{\rm L,sa}/E_0\) at the end of autoresonance, shown by the blue curve. The two points indicate representative cases with laser field amplitudes \(a = 0.1\) (left) and \(a = 0.2\) (right)}.}
\label{scaling} 
\end{figure}

The numerical results characterizing the plasma wave properties for linearly varying density profiles are shown in Fig.~\ref{90_180_360linear}.  In the simulations, the laser amplitudes are set to $a_{1,2}$ = 0.1. The plasma density gradient length is varied across the panels: $L_{\rm gra} = 180\pi$ in the left column, $360\pi$ in the middle, and $720\pi$ in the right. The top panels of Fig.~\ref{90_180_360linear} show the spatial profiles of the plasma wave at multiple time snapshots, illustrating the evolution and amplitude growth under different gradient conditions. The bottom panels display the corresponding time evolution of the local phase difference $\Phi(x)$ between the driving beat and the plasma wave, defined as $\Phi(x) = \phi_1(x) - \phi_2(x) - \phi_L(x)$,
where $\phi_{1,2,L}$ denote the complex phases of the two laser beams and the plasma wave, respectively. Resonant energy transfer is achieved when the three-wave interaction remains phase-matched, corresponding to $\cos(\Phi) \approx 1$~\cite{lindbergpop,lindbergprl}.

\begin{figure*}[htbp]
    \centering
    \includegraphics[width=1\linewidth]{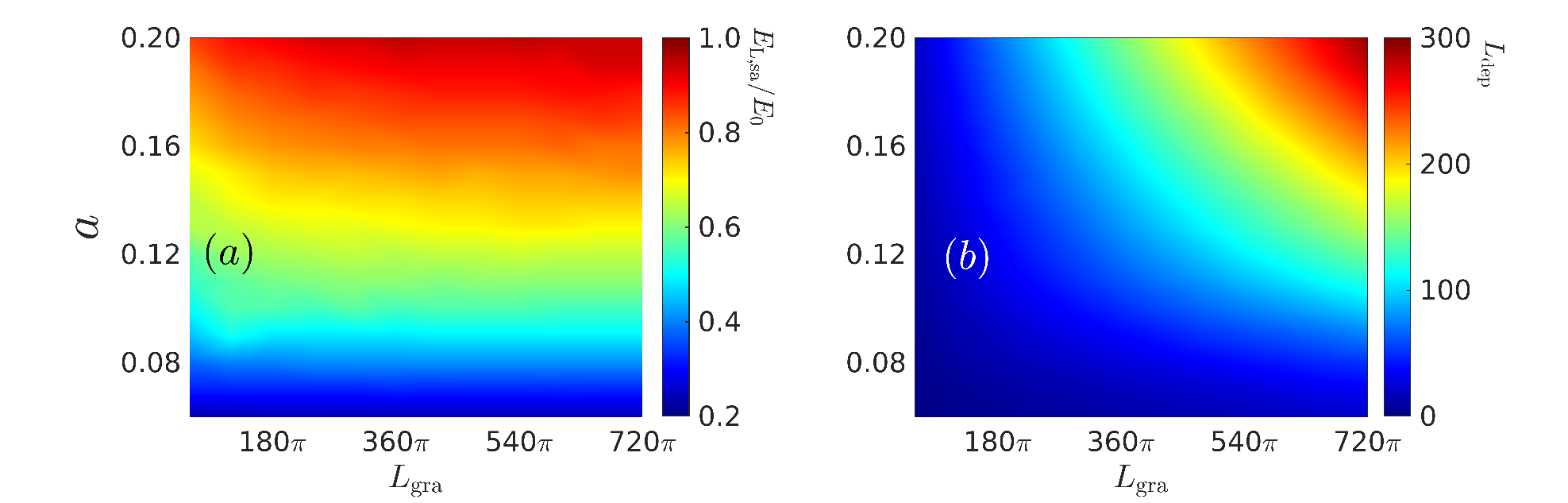}
    \caption{Autoresonant plasma wave excitation in a linearly increasing plasma density profile. The saturated plasma wave amplitude (a) and the dephasing length in (b) are shown as functions of the density gradient length and laser amplitude, for the reference plasma density $n_{\rm re}(x_{\rm re})=0.0004n_{\rm cr}$.}  
\label{scan-intensity} 
\end{figure*}

As shown in the top panels, increasing the density gradient length broadens the spatial extent of the resonant region. In this regime, the plasma wave amplitude exceeds the RL limit. For reference, the dotted black curve indicates the RL limit at the reference point, given by $E_{\rm RL}(x=x_{\rm re})/E_0 =(16 a_1 a_2 / 3)^{1/3}$, where $E_0$ is the wave-breaking field at $x_{\rm re}$.

Equation~\eqref{el} (solid black curve) reproduces well the growth of the wavefront, confirming that the density-induced phase shift $\delta k_{\rm s}$ compensates the nonlinear shift $\delta k_{\rm non}$, thereby sustaining autoresonant excitation. 
The phase-locked region, coinciding with the resonant zone, systematically broadens with increasing $L_{\rm gra}$, as shown in the lower panels. The spatio-temporal trajectory of the maximum plasma-wave amplitude (solid black line) closely follows the rear boundary of this expanding phase-locked region.

To illustrate the robustness of autoresonant plasma wave excitation, Fig.~\ref{a0.10.2}(a) shows the maximum value of $E_L/E_0$ over the simulation domain as a function of time for a wide range of gradient lengths, with fixed laser amplitudes $a_{1,2}$= 0.1. Despite the variation in gradient scale, all cases reach the same final amplitude by the end of the autoresonant interaction, indicating that the process is highly resilient to changes in $L_{\rm gra}$. Once the autoresonance terminates, the peak amplitude of the plasma wave remains localized, at the \emph{turning point}, $x_t$, at which the plasma wave amplification ceases. This localized structure undergoes fast oscillation in time. However, the time required to reach saturation increases with the gradient length; equivalently, the plasma wave must propagate a longer distance before amplification ceases. The horizontal dotted line denotes the RL limit evaluated at the reference point. 

Based on the compensation between these two wavenumber shifts, we can estimate the dephasing length ($L_{\rm dep}=x_t-x_{\rm re}$) as, 
\textcolor{black}{
\begin{equation}
     (L_{\rm dep}/L_{\rm gra})^{i}=n_e|_{\rm sa}/n_{\rm re}-1.  \label{linear dephase1}
\end{equation}}
With $i=1$ corresponding to the linear density gradient, \textcolor{black}{and $n_e|_{\rm sa}$ the plasma density at the end of the autoresonance. The relation between the $n_e|_{\rm sa}/n_{\rm re}$ and the electric field $E_{\rm L, sa}/E_0$ at the end of the autoresonance is given by solving Eq.~\eqref{el}, and the solution is plotted by the blue curve in Fig.~\ref{scaling}.} Eq.~\eqref{linear dephase1} indicates that the dephasing length $L_{\rm dep}$ scales linearly with the plasma density gradient length $L_{\rm gra}$, since the term in square brackets remains approximately constant for fixed laser amplitude. This constancy arises because the saturated plasma wave amplitude $E_{\rm L,sa}$ at the end of the autoresonant interaction is nearly identical across different gradients, as shown in Fig.~\ref{a0.10.2}(a).

Figure~\ref{a0.10.2}(c) presents the measured dephasing lengths $L_{\rm dep}$ from the simulations (red squares) as a function of $L_{\rm gra}$. \textcolor{black}{Taking the saturated plasma wave amplitude \(E_L/E_0 \approx 0.6\), as observed in Fig.~\ref{a0.10.2}(a), the corresponding density is \(n_e|_{\rm sa}/n_{\rm re} \simeq 1.06\), as indicated by the blue curve in Fig.~\ref{scaling}. Substituting this value into Eq.~\eqref{linear dephase1} yields a prediction that shows very good agreement with the numerical results, as shown by the green line in Figure~\ref{a0.10.2}(c).}


%
\begin{figure}[htbp]
    \centering
    \includegraphics[width=1\linewidth]{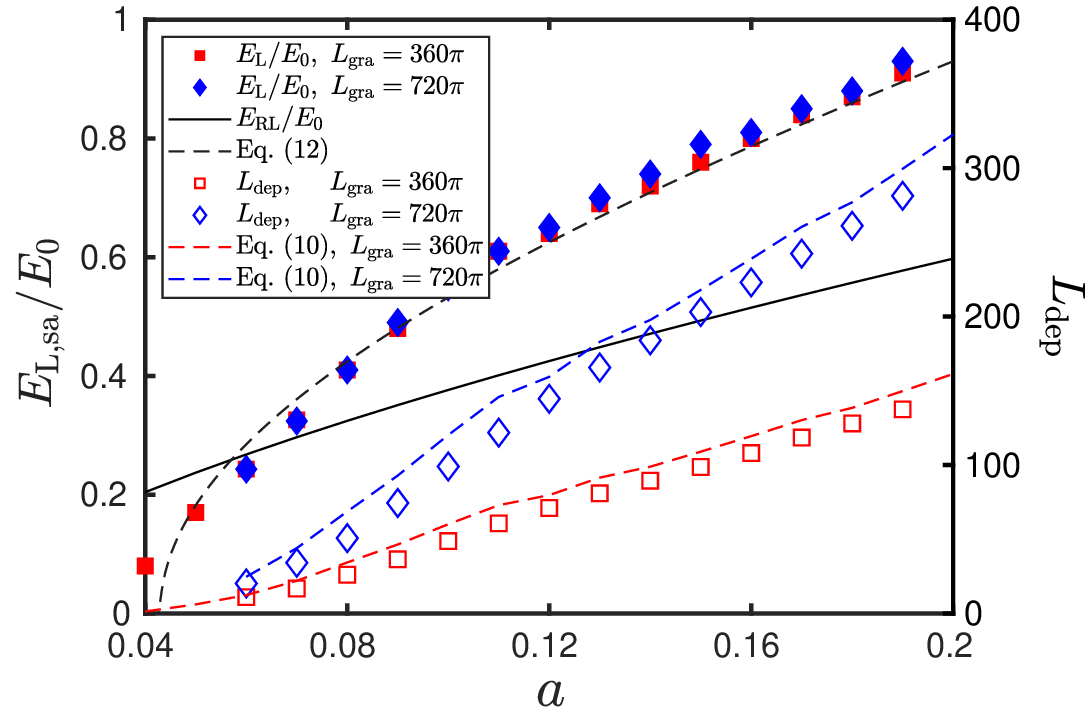}
    \caption{Saturated plasma wave amplitude $E_{\rm L,sa}/E_0$ (left axis) as a function of laser amplitude $a$ for the linear density profiles with $L_{\rm gra} = 360\pi$ (solid red squares) and $L_{\rm gra} = 720\pi$ (solid blue diamonds); The solid black curve shows the RL limit, while the dashed black curve shows the data-driven estimate from Eq.~\eqref{linear intensity}. Corresponding dephasing lengths $L_{\rm dep}$ (right axis) are plotted using open symbols for both cases. The dashed red and blue curves indicate the analytical predictions from Eq.~\eqref{linear dephase1}.}
\label{dephasinglength} 
\end{figure}

Figures~\ref{a0.10.2}(b) and \ref{a0.10.2}(d) show the plasma wave evolution for higher laser intensities, $a_{1,2}$= 0.2. Similar to the lower-amplitude case shown in Fig.~\ref{a0.10.2}(a), the plasma wave amplitude grows beyond the RL limit through the autoresonant plasma wave excitation mechanism and saturates at a comparable level (slightly below the nonrelativistic wave-breaking limit) across different gradient lengths. In Fig.~\ref{a0.10.2}(d), the dephasing lengths $L_{\rm dep}$ extracted from simulations are plotted as blue diamonds against $L_{\rm gra}$. Once again, the linear trend is well captured by Eq.~\eqref{linear dephase1}, as shown by the green line, when the normalized wave amplitude is taken as \(E_L/E_0 \approx 0.95\), consistent with the saturation value observed in Fig.~\ref{a0.10.2}(b), \textcolor{black}{which leads to \(n_e|_{\rm sa}/n_{\rm re} \simeq 1.14\) (shown by the blue curve in Fig.~\ref{scaling}).}

We highlight two representative cases, marked by open black circles in Figs.~\ref{a0.10.2}(c) and \ref{a0.10.2}(d), which share the same density gradient length $L_{\rm gra} = 1440\pi$ but differ in laser amplitude. The corresponding phase evolution for these two cases is shown in the inset of Fig.~\ref{a0.10.2}(c). We can indeed observe that a higher laser amplitude results in the broader resonant regime, i.e., a longer dephasing length $L_{\rm dep}$. This can be correlated to higher values of $E_L$ observed further away from $x_{\rm re}$ at higher amplitude, as we will discuss further. 

\begin{figure*}[htbp]
    \centering
    \includegraphics[width=1\linewidth]{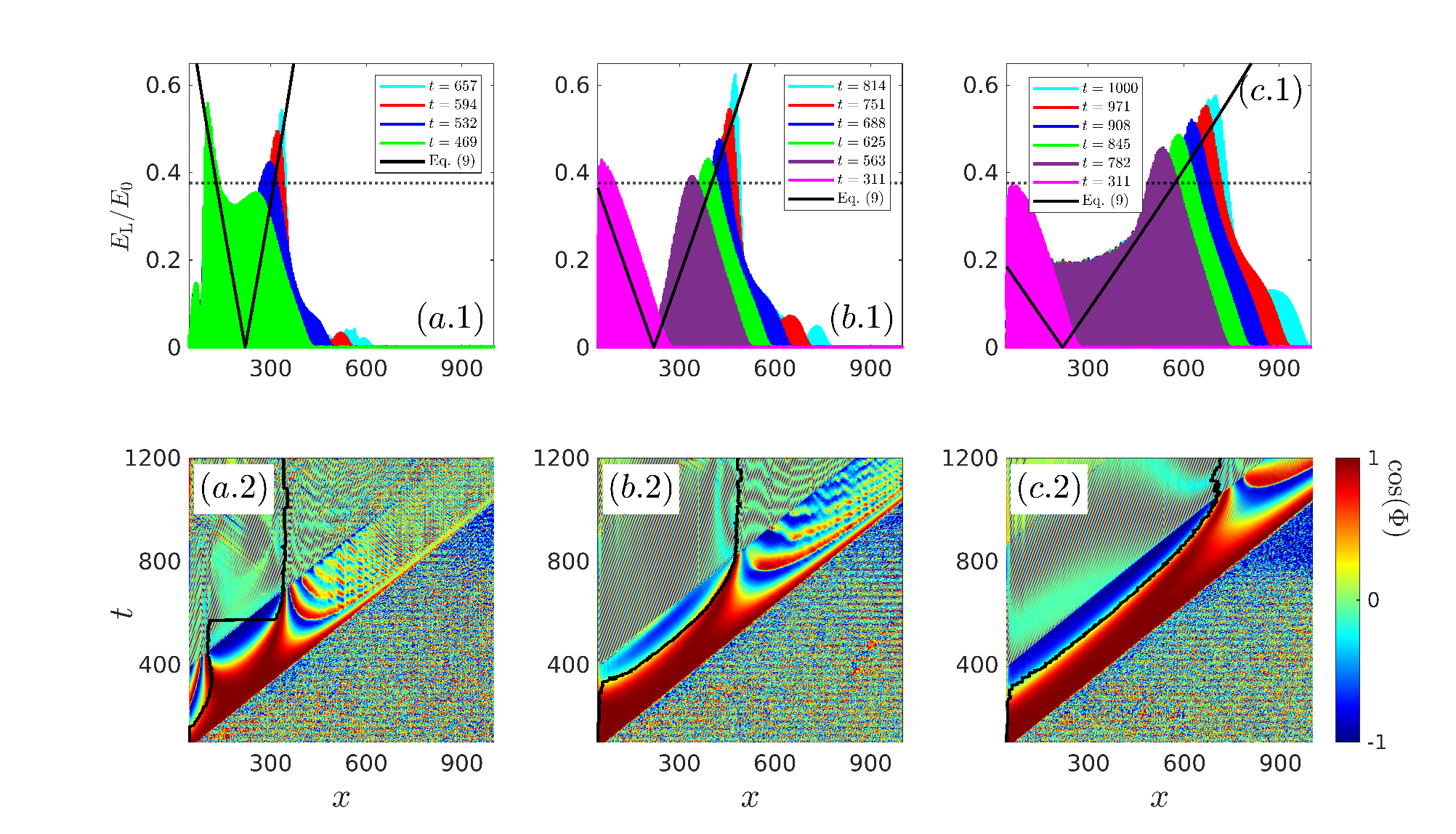}
    \caption{Top panels: The spatial profile of the plasma wave at different times, for parabolic density profiles. Solid black curves correspond to the amplitude predictions from Eq.~(\ref{el}), and the dotted black line indicates the RL limit at the reference point. Bottom panels: The spatio-temporal evolution of the phase difference between the three interacting waves. Here, the solid black curves mark the location of the highest plasma wave amplitude. The density gradient length is increased between the panels from left to right: $L=180\pi$, $L=360\pi$, and $L=720\pi$. The simulations use the normalized laser amplitude $a_{1,2}$=0.1 and the reference plasma density $n_{\rm re}(x_{\rm re})=0.0004n_{\rm cr}$.} 
    \label{parabolic_90_180_360}
\end{figure*}

To systematically explore the dependence of autoresonant dynamics on laser and plasma parameters, we perform a parameter scan with laser amplitudes ranging from $a_1 = a_2 = 0.05$ to $0.2$, and gradient lengths $L_{\rm gra}$ varying from $60\pi$ to $720\pi$. The resulting saturated plasma wave amplitudes $E_{\rm L,sa}/E_0$ and the corresponding dephasing lengths $L_{\rm dep}$ are presented in Figs.~\ref{scan-intensity}(a) and \ref{scan-intensity}(b), respectively. As expected, Fig.~\ref{scan-intensity}(a) shows that for a fixed laser amplitude, the final amplitude $E_{\rm L,sa}/E_0$ is nearly independent of $L_{\rm gra}$, with only a slight reduction observed for shorter gradient lengths.  Fig.~\ref{scan-intensity}(b) instead  shows  that $L_{\rm dep}$ increases with both the density gradient and the driving laser amplitude. The linear dependence of $L_{\rm dep}$ on $L_{\rm gra}$, already demonstrated in Figs.~\ref{a0.10.2}(c) and \ref{a0.10.2}(d), is consistently observed across the full parameter range.

In Fig.~\ref{dephasinglength} we show the normalized electric field amplitude $E_{\rm L,sa}/E_0$ as a function of the laser amplitude $a$. The red solid squares correspond to simulations with $L_{\rm gra} = 360\pi$, while the blue solid diamonds represent $L_{\rm gra} = 720\pi$. For reference, the black curve shows the RL limit as a function of $a$. Notably, for laser intensities $a > 0.06$, the autoresonant plasma wave exceeds the RL limit, confirming the efficacy of autoresonant high-amplitude plasma wave excitation with spatial density gradients.

On the right vertical axis of Fig.~\ref{dephasinglength}, the corresponding dephasing lengths $L_{\rm dep}$ are plotted as open symbols for both gradient cases. \textcolor{black}{By inserting the observed $E_{\rm L,sa}/E_0$ values into Eq.~\eqref{el}, then the obtained $n_e|_{\rm sa}/n_{\rm re}$ is further input into Eq.~\eqref{linear dephase1}}, the calculated dephasing lengths are shown as dashed red and blue curves, demonstrating excellent agreement with simulation data. Both numerical and analytical results exhibit a clear linear dependence of $L_{\rm dep}$ on laser amplitude, i.e., $L_{\rm dep} \propto a$. This scaling is consistent with the gain saturation behavior in stimulated Raman scattering~\cite{RamanGain}, where $L_{\rm dep} \sim a L_{\rm gra}$ in the linear density gradient. 

The relationship between laser amplitude and plasma wave amplitude in this regime is not trivial. However, we note that increasing the laser field from $a_{1,2}$= 0.1 to 0.2 leads to a linear increase in the dephasing length from $L_{\rm dep}$$\approx$110 to 310, for a fixed gradient length $L_{\rm gra} = 720\pi$. The dephasing length is found to be well approximated by a linear dependence on the laser amplitude
\textcolor{black}{\begin{equation}
    f(a) = L_{\rm dep}(a)/720\pi\approx (200a-9)/72\pi. \label{linear relation}
\end{equation}}
Inserting this into Eq.~(\ref{el}) by substituting $n_e(x)/n_{\rm re}$ with $\mathcal{F}(a)=1+f(a)$, we obtain an empirical expression for the plasma wave amplitude,
\textcolor{black}{
\begin{equation}
    \left(\frac{E_L}{E_0}\right)^2 = \frac{16}{3}\mathcal{F}(a)\left(\frac{\sqrt{\mathcal{F}(a)}}{2-\sqrt{\mathcal{F}(a)}}-1\right). \label{linear intensity}
\end{equation}}
The predicted plasma wave amplitudes from Eq.~\eqref{linear intensity} are shown as the black dashed line in Fig.~\ref{dephasinglength}, exhibiting excellent agreement with simulation results. Together, Eqs.~\eqref{linear dephase1} and \eqref{linear intensity} describe the autoresonant plasma wave excitation process in a linearly increasing plasma density profile.

\subsection{Autoresonant plasma wave excitation in a parabolic density profile \label{parabolic part1}}

In this subsection, we describe the plasma wave excitation process in a parabolic density profile. The simulation setup to get Fig.~\ref{parabolic_90_180_360}, apart from the density profile being parabolic instead of linear, is analogous to Fig.~\ref{90_180_360linear}. Namely, the laser amplitudes are $a_{1,2}$= 0.1, and the plasma density gradient length is varied across the panels; the top panels illustrate the evolution and amplitude growth under different gradient conditions and the bottom panels show the corresponding time evolution of the local phase difference $\Phi(x)$ between the driving beat and the plasma wave. Efficient energy transfer occurs when the three-wave coupling maintains phase matching, i.e., $\cos\Phi \simeq 1$.

In Fig.~\ref{parabolic_90_180_360}(a.1), with $L_{\rm gra} = 180\pi$, at the early stage ($t \leq 470$), as the plasma wave is driven from left to right, it dynamically senses the plasma-frequency gradient and adapts its structure to maintain phase-locking with the driving beat. This mechanism enables resonant amplification before the nominal linear reference point $x_{\rm re}$ is reached. Beyond this point, as the wave continues to propagate through $x_{\rm re}$, the tailored plasma density triggers autoresonance: the wave, initially in linear resonance, remains amplified as the density-induced wavenumber shift is effectively compensated by the nonlinear wavenumber shift, thereby sustaining phase-locking.  Despite the different amplification mechanisms, self-organization (left) and autoresonance (right), the plasma wave amplitude exhibits a symmetric structure owing to the symmetric density profiles, and is well described by Eq.~(\ref{el}), as shown by the solid black line. The dotted black line indicates the RL limit at the reference point $E_{\rm RL}(x=x_{\rm re})/E_0 =(16 a_1 a_2 /3)^{1/3}$, where $E_0$ is the wave-breaking field at the reference point $x_{\rm re}$. Both mechanisms, self-organization and autoresonance, enable amplification beyond the linear RL limit.

\begin{figure}[htbp]
    \centering
    \includegraphics[width=1\linewidth]{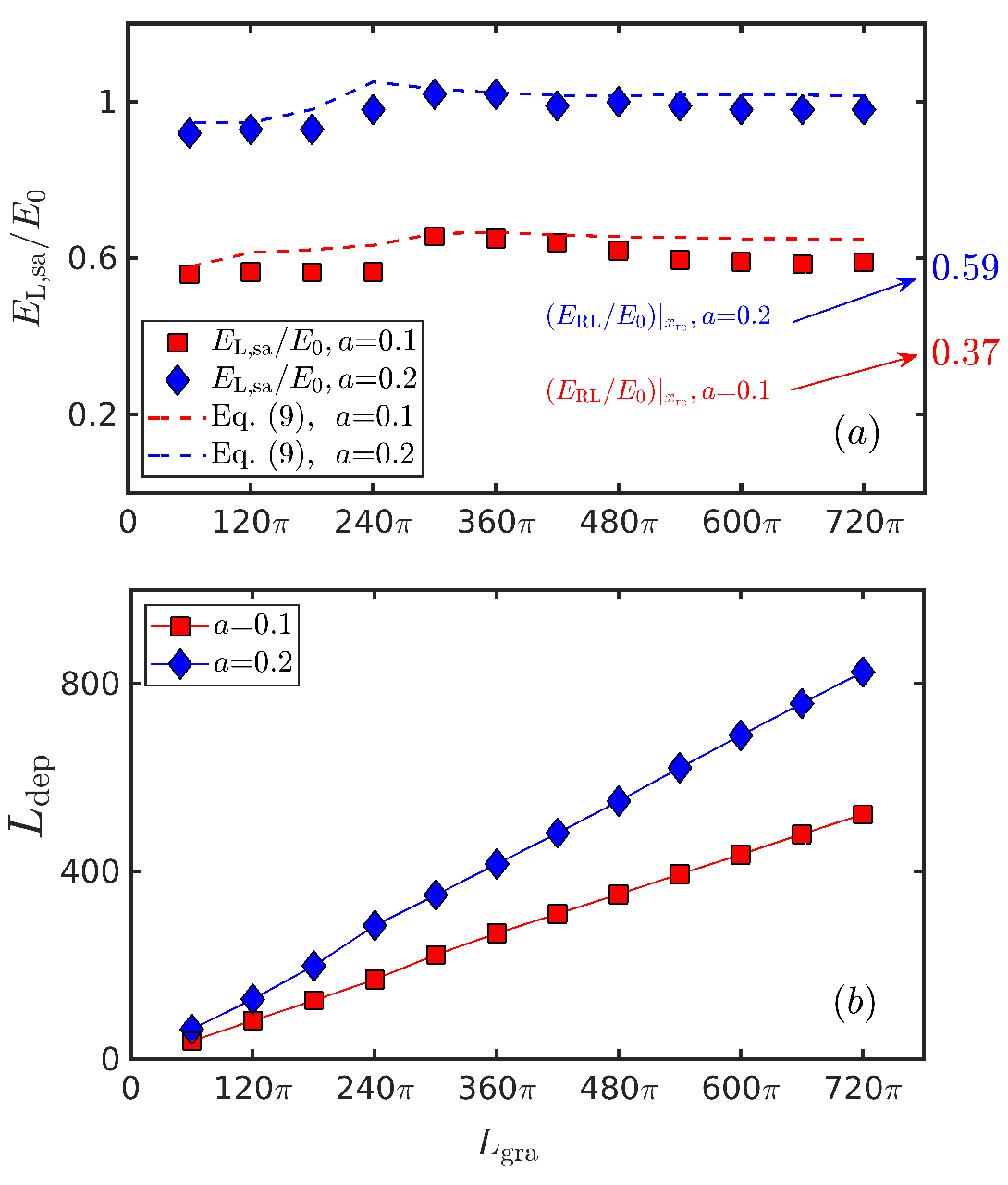}
    \caption{Autoresonant plasma wave excitation in a parabolic density profile for varying gradient lengths and laser intensities. Saturated electric field amplitude $E_{\rm L,sa}/E_0$ as a function of the density gradient length $L_{\rm gra}$ for laser amplitudes $a_{1,2}$=0.1 (red squares) and 0.2 (blue diamonds) shown in (a). For reference, the red and blue arrows indicate the RL limits at the reference point. The two dashed colored curves correspond to the calculated values of Eq.~(\ref{el}) by taking the observed dephasing length $L_{\rm dep}$ shown in (b).}
    \label{0.10.2parabolic} 
\end{figure}

The phase information is provided in Fig.~\ref{parabolic_90_180_360}(a.2), which shows the spatio-temporal evolution of the phase difference $\Phi$. Two distinct phase-locking regions appear symmetrically on either side of the resonant region. The black trajectory, corresponding to the location of peak plasma wave amplitude, is initially localized on the left side and shows a slight back-propagation of the peak plasma wave amplitude before it progressively transitions to the right.  

\begin{figure*}[htbp]
    \centering
   \includegraphics[width=1\linewidth]{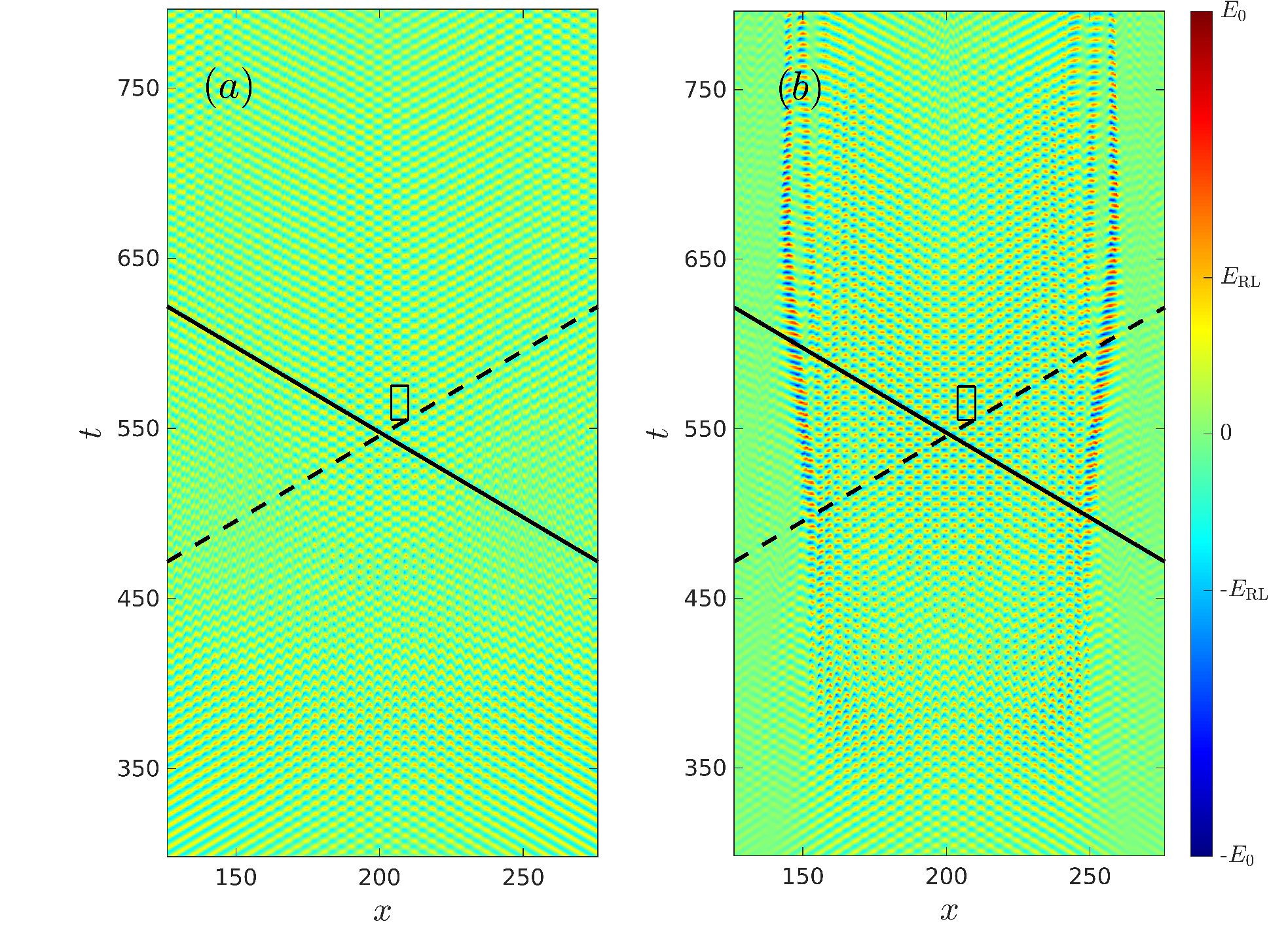}
    \caption{The spatio-temporal structure of the excited plasma wave pattern is shown, (a) in a spatially homogeneous plasma, and (b) in a parabolic density profile with $L_{\rm gra}=60\pi$. A pair of lasers beams is injected from the left boundary, with their rear indicated by the dashed line, and another pair of laser beams is injected from the right boundary, with their rear indicated by the solid line. The four laser beams have the same amplitude $a_{1,2,3,4}=0.1$, and the linear polarizations of the beams within each pair are parallel, while the two pairs have orthogonal polarization. In the homogeneous plasma the density is $n_{\rm e}=0.0004n_{\rm cr}$ throughout the domain, while for the parabolic profile this value is taken at the linear resonant point. The black rectangles delineate the regions shown in Fig.~\ref{small-scale}.}  
\label{gratting} 
\end{figure*}

The gradient length is increased to $L_{\rm gra} = 360\pi$ in Fig.~\ref{parabolic_90_180_360}(b) and $L_{\rm gra} = 720\pi$ in Fig.~\ref{parabolic_90_180_360}(c), resulting in the broader phase-locking regime (bottom panels), then the longer resonant growth length (top panels). While for $L_{\rm gra} = 720\pi$, the spatial extent on the left side of the profile, specifically, the region from the simulation boundary to the onset of resonance at $x_{\rm re}$, is too narrow to support self-organization. This limitation is evident in Fig.~\ref{parabolic_90_180_360}(c.1), where no significant amplification develops in the left wing. In this case Eq.~(\ref{el}) fails to describe the envelope, and the amplitude remains at the RL level. Similarly to what observed in the linear density profile, Fig.~\ref{parabolic_90_180_360} shows that, while the peak amplitude of the wave is roughly independent of the gradient length, the spatial excursion of the wave and the time needed to reach the peak value increase with the gradient length. As a consequence, for steep gradients, a shorter laser duration than the one used in this study ($ T_0 = 100 \pi/\omega_{\rm re}$) can be enough.  

To further characterize the waves in the parabolic density profiles, we performed simulations scanning the gradient length $L_{\rm gra}$ from $60\pi$ to $720\pi$ for two fixed laser amplitudes: $a_{1,2}$= 0.1 and 0.2. The corresponding saturated plasma wave amplitudes $E_{\rm L,sa}$, located on the right wing of the plasma, are shown in Fig.~\ref{0.10.2parabolic}(a). Red squares and blue diamonds denote the results for $a_{1,2}$=0.1 and 0.2, respectively. For reference, the red and blue arrows indicate the RL limits at the reference point. In Fig.~\ref{0.10.2parabolic}(b), we plot the dephasing length $L_{\rm dep}=x_t-x_{\rm re}$ for different gradient lengths $L_{\rm gra}$ at laser amplitude $a_{1.2}$=0.1 (red squares) and 0.2 (blue diamonds). We recall that the turning point $x_t$ marks the point where plasma wave amplification ceases and saturation sets in. Similarly to the linear density gradient case, we find that the dephasing length is proportional to the gradient length for a given laser amplitude. The proportionality between $L_{\rm dep}$ and $L_{\rm grad}$ is consistent with the fact that the saturated electric field $E_{\rm L,sa}/E_0$ is roughly constant for a given laser amplitude (see Fig.~\ref{0.10.2parabolic}(a)). Inserting the observed dephasing length $L_{\rm dep}$ into Eq.~(\ref{el}), The resulting values of $E_{\rm L,sa}/E_0$ show excellent agreement, see the two dashed colored curves in Fig.~\ref{0.10.2parabolic}(a). 

\section{Two-phase nonlinear electron plasma waves \label{crystal}}

From Sec.~\ref{parabolic-part}, we observe that the plasma-wave structure can be manipulated through tailored density profiles, either by self-organization or via autoresonance. This capability highlights the potential of plasmas to function as optical elements through controlled shaping of their refractive-index profiles. In this section, we further explore the extent of plasma configurability by demonstrating a scheme for the realization of a plasma quasicrystal. 

The simulation configuration used to model the plasma quasicrystal is outlined as follows. In contrast to the setup described in Sec.~\ref{parabolic part1}, a second pair of laser beams is injected from the right, generating an additional ponderomotive force. 
The lasers entering from the right, with amplitudes $a_3$ and $a_4$, are polarized perpendicular to the polarization of the lasers injected from the left. Both pairs of lasers are otherwise (aside from propagation direction and polarization) identical and not chirped. The interference of the two driving pairs produces a two-phase–like structure within the parabolic density channel. To compare with the parabolic density results, we considered the same set-up in a homogeneous plasma. Fig.~\ref{gratting} presents the spatio-temporal evolution of the plasma wave driven in the homogeneous plasma $(a)$ and parabolic plasma profile $(b)$. For reference, the solid (dashed) line marks the rear boundary of the lasers incident from the right (left) side.   
\begin{figure}[htbp]
    \centering
    \includegraphics[width=1\linewidth]{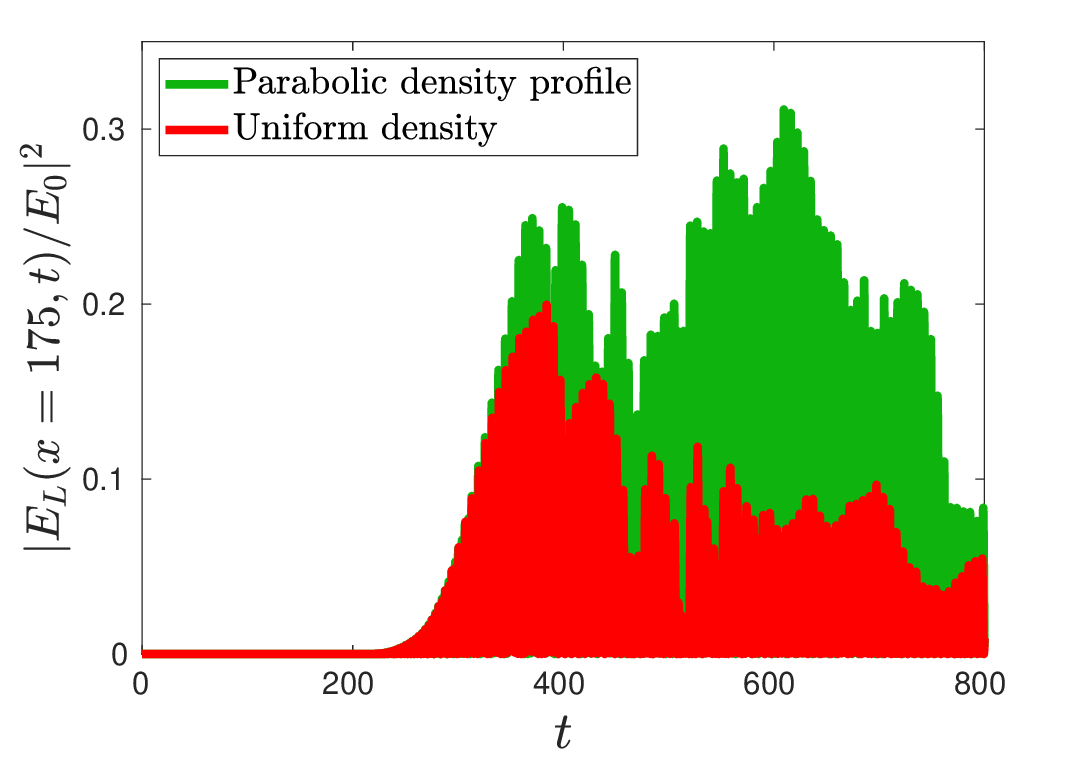}
    \caption{\textcolor{black}{Temporal evolution of the normalized plasma wave energy \( |E_L/E_0|^2 \) at the fixed position \( x = 175 \). 
    The green (red) regions correspond to parabolic (uniform) plasma densities
    The laser pulses exit this position at approximately \( t \approx 560 \).}}  
\label{energy_decay} 
\end{figure}

In Fig.~\ref{gratting}(a), with a homogeneous plasma, the plasma wave is primarily excited within the overlap region of the two laser pairs, \textcolor{black}{and the plasma wave amplitude saturates at the RL limit due to nonlinear detuning, although it shows the two-phase electron plasma wave structure, i.e., a crystal-like quasiperiodic spatiotemporal structure~\cite{Munirovpre,Munirovprr}.} 
The plasma wave decays to lower amplitude after the lasers cease to overlap. 
Instead, in Fig.~\ref{gratting}(b), where a parabolic plasma density profile is used, not only is the plasma wave excited to much higher values of the amplitude within the overlap region, above the RL limit, but it also persists after the lasers have left the interaction region. \textcolor{black}{This is explicitly shown in Fig.~\ref{energy_decay}, where the temporal evolution of the plasma wave energy \( |E_L/E_0|^2 \) at a fixed position \( x = 175 \) is plotted. At this location, the laser pulses exit the region at approximately \( t \approx 560 \). In the uniform density scenario (red curve), the driven plasma wave decays to a lower amplitude after the lasers exit. By contrast, the parabolic density profile (green curve) allows the plasma wave amplitude to remain nearly constant. Notably, there exists a time interval during which the plasma wave continues to grow following laser exit.} The spatial extension along $x$ of the region where the wave develops is related to the dephasing length.
In Fig.~\ref{gratting}(b), the plasma wave structure is stable over a period of approximately $t_{\rm g}\approx 100$ (approximately from $t=550$ to $650$). Assuming an $800\,\rm nm$ laser, a reference density of $n_e \approx 7 \times 10^{17}\ \rm cm^{-3}$, \textcolor{black}{a plasma density gradient length of \(1.2\,\mathrm{mm}\), and a total plasma length of \(2\,\mathrm{mm}\), with the reference point located at the plasma center}, this corresponds to a duration of $2.1\,\rm ps$. This structure duration is sufficient for an effective manipulation of short laser pulses.

\begin{figure}[htbp]
    \centering
    \includegraphics[width=1\linewidth]{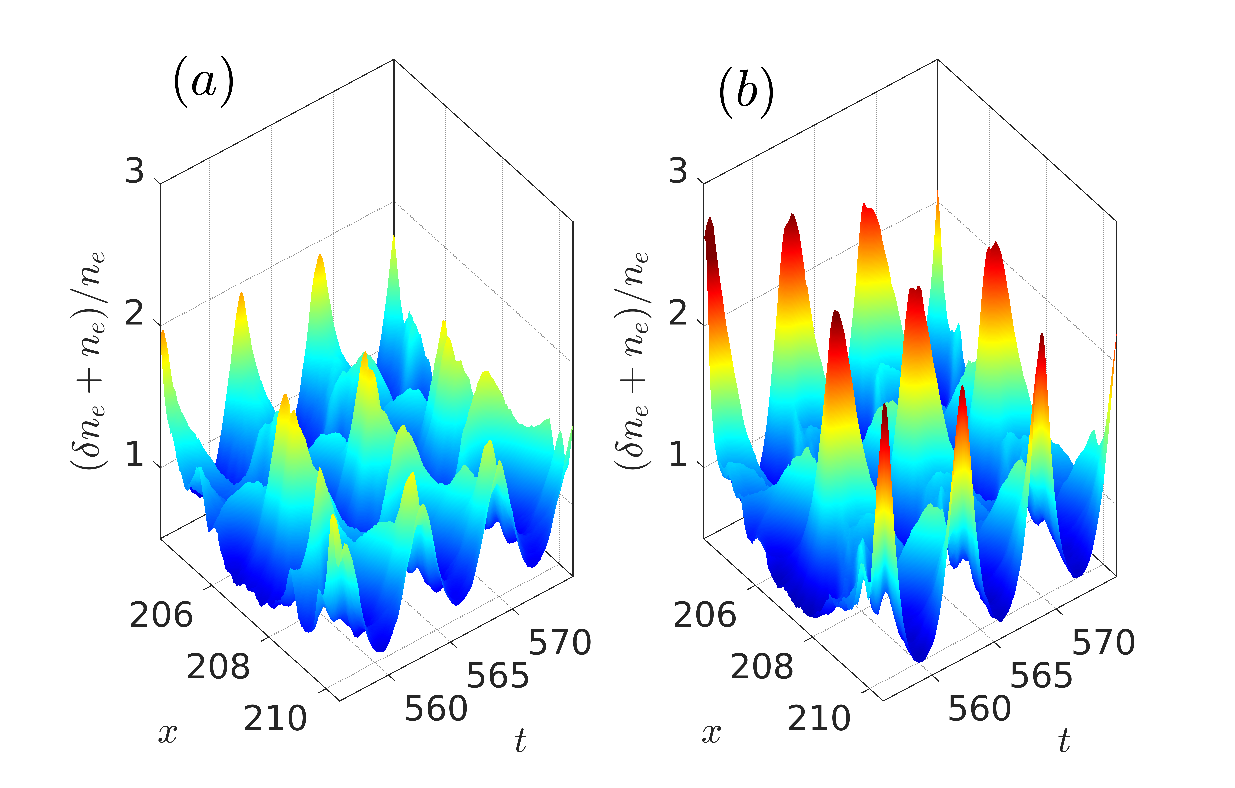}
    \caption{Zoom-in of the normalized perturbed plasma density evolution, $(\delta n_e+n_e)/n_e$, where $n_e$ and $\delta n_e$ are the background and the perturbed density respectively. (a): homogeneous plasma, (b): parabolic density profile; these correspond to regimes indicated by the black rectangles in the left and right panels of Fig.~\ref{gratting}.}  
\label{small-scale} 
\end{figure}

Figure~\ref{small-scale} shows the plasma density structure as $(\delta n_e+n_e)/n_e$, where $\delta n_e$ represents the density perturbation caused by the plasma wave, and $n_e$ is the background plasma density. The plot shows a zoom of the plasma density structure within a small spatio-temporal region highlighted by a box in  Fig.~\ref{gratting}. This region was selected because the lasers had already exited the nominal linear reference point. The comparison between Fig.~\ref{small-scale}(a) and (b) underscores the impact of self-organization and autoresonance on the plasma waves. Both mechanisms enhance the amplitude of the plasma wave and extend its lifetime after the ponderomotive forcing is not present anymore (since the lasers have exited the region). These simulations indicate that the two-phase spatio-temporal structure investigated in a fluid model in Ref.~\onlinecite{Munirovpre} using chirped lasers can be created with a kinetic PIC simulation with a parabolic density profile and unchirped lasers.

\section{CONCLUSIONS \label{conclusion}}

We have investigated the autoresonance of plasma beat-wave excitation in tailored plasma density configurations using fully kinetic PIC simulations. Our results demonstrate that the nonlinear wavelength shift of the beat-driven plasma wave, responsible for the saturation at the RL limit in conventional PBWA schemes~\cite{Rosenbluthprl1972}, can be effectively compensated for by a spatial variation of the plasma frequency. This compensation enables continuous phase-locking between the driving laser fields and the plasma wave, allowing the wave amplitude to exceed the RL limit and, in some cases, approach the nonrelativistic wave-breaking limit. We show that the saturated amplitude of the autoresonant plasma wave remains remarkably insensitive to the gradient length, across both linearly increasing and parabolic density profiles. These findings offer new perspectives for applications in high-power terahertz generation via linear mode conversion in plasma gradients and for designing structured plasma waves by engineering the background density. The demonstrated control over phase-locking and amplitude growth may become a new pathway for tunable plasma-based photonic devices. \textcolor{black}{Meanwhile, the cold plasma approximation is adopted in this work by setting the electron temperature to \( T_e \approx 0.01\,\mathrm{keV} \). At higher electron temperatures, kinetic effects would reduce the plasma wave wavebreaking limit and modify the plasma wave dispersion relation, leading to a frequency shift that may detune the resonant condition unless the laser frequency difference is adjusted accordingly. Ion motion is expected to be negligible for the laser and plasma conditions considered here, consistent with our previous studies~\cite{luoprr}.}

As discussed in Refs.~\onlinecite{luoprr,luojpp}, a highly underdense plasma with $n_e/n_{\rm cr} \sim 0.0004$ was chosen to suppress fluid nonlinearities in the laser propagation and kinetic nonlinearities in the plasma wave response. For experimental relevance, this corresponds to an electron density of $n_e \approx 7 \times 10^{17}\ \rm cm^{-3}$ when using a Ti:sapphire CPA laser at $800\ \rm nm$ wavelength. While our simulations focused on this underdense regime, the autoresonance mechanism is expected to remain effective at higher densities, since it does not require long-distance plasma wave propagation as in conventional plasma-based acceleration schemes~\cite{luoprr}.
The autoresonant excitation via autoresonance can generate plasma waves reaching the wave-breaking amplitude with a moderate laser amplitude of $8.5 \times 10^{16}\ \rm W/cm^2$, and importantly, without the need for frequency chirping. Although the laser pulse duration employed in our study was $T_0 = 100\pi/\omega_{\rm pe}$ (corresponding to 6.7 ps), such durations are not strictly necessary in practice. For example, with a density gradient length of $L_{\rm gra} = 3.6\ \rm mm$ (i.e., $L_{\rm gra}k_{\rm pe} = 180\pi$), the effective interaction time is around 3 ps, as seen in Figs.~\ref{90_180_360linear}(a) and~\ref{parabolic_90_180_360}(a). The required laser duration could be even shorter, particularly when considering experimentally observed density ramps, such as those reported in Ref.~\onlinecite{linearmode5}.

A two-phase quasicrystalline-like structure was excited using two pairs of counter-propagating lasers, suggesting new possibilities for applications in plasma photonics. This study extended prior work by demonstrating that the structures exist in kinetic one-dimensional models. However, higher-dimensional effects such as wave-front bowing of the plasma wave and density perturbation filamentation~\cite{luojpp} that appear at a late stage of the interaction, could degrade the desired optical performance of a structure based on this concept.  This motivates follow-up studies of autoresonance in multi-dimensional geometry, which may involve a limitation on the driving pulse duration.

As previously demonstrated in Ref.~\onlinecite{linearmode3}, high-power terahertz (THz) radiation can be generated via linear mode conversion of plasma waves in inhomogeneous plasmas, particularly in positive density gradients. The autoresonance of plasma beat waves presented here offers a promising method for precise control over plasma wave properties and thus opens a potential pathway toward tunable THz emission. However, exploration of this application lies beyond the scope of the present work.

\vspace*{-.3cm}\begin{acknowledgements}\vspace{-.35cm}
This project received funding from the Knut and Alice Wallenberg Foundation (Grant No.~KAW 2020.0111 and 2023.0249). The computations were enabled by resources provided by the National Academic Infrastructure for Supercomputing in Sweden (NAISS), partially funded by the Swedish Research Council through grant agreement No.~2022-06725 and 2021-03943. C.~Riconda and J.~S.~Wurtele thank the Berkeley-France Fund for support of this research.
\end{acknowledgements}

\vspace*{-.6cm}
\section*{Author declarations}\vspace{-.45cm}
\subsection*{Conflict of Interest}\vspace{-.45cm}
The authors have no conflicts to disclose.
\vspace*{-.6cm}
\section*{Data Availability}\vspace{-.45cm}
The data that support the findings of this study are available from the corresponding author upon reasonable request.
\section*{references}\vspace{-.35cm}
\bibliography{reference.bib}
\end{document}